\documentclass[12pt]{article}

\usepackage[margin=1in]{geometry}
\usepackage{amsfonts, amsmath, amssymb}
\usepackage[none]{hyphenat}
\usepackage{fancyhdr}
\usepackage{float}
\usepackage{framed} 

\usepackage{array}
\usepackage{lscape} 

\usepackage{listings} 
\usepackage{longtable}

\usepackage{url}
\usepackage{rotating}

\usepackage{natbib}

\begin{document}

\begin{titlepage}
\begin{center}

\vspace*{1cm}
\Large{\textbf{}}\\
\vfill
\line(1,0){450}\\
\Large{\textbf{Prediction intervals for overdispersed binomial endpoints and their 
application to toxicological historical control data}}\\

\line(1,0){450}\\
\vfill

\flushleft
\small
Max Menssen$^{1}$, Jonathan Rathjens$^2$\\

1: menssen@cell.uni-hannover.de, Department of Biostatistics, Leibniz University Hannover \\
2: jonathan.rathjens@chrestos.de, Early Development Statistics, Chrestos GmbH, Essen \\
Corresponding author: Max Menssen\\

\textbf{Data Availability Statement}\\
The data that support the findings of this study are available on request from the 
corresponding author. The data are not publicly available due to privacy or ethical 
restrictions.\\

\textbf{Conflict of interest}\\
No, there is no conflict of interest.

\end{center}
\end{titlepage}

\setcounter{page}{1}


\section*{Abstract}
For toxicology studies, the validation of the concurrent control group by historical control data (HCD)
has become requirements. This validation is usually done by historical control limits (HCL),
which should cover the observations of the concurrent control with a predefined level of confidence.
In many applications, HCL are applied to dichotomous data, e.g. the
number of rats with a tumor vs. the number of rats without a tumor (carcinogenicity studies)
or the number of cells with a micronucleus out of a total number of cells. Dichotomous
HCD may be overdispersed and can be heavily right- (or left-) skewed, which is usually not
taken into account in the practical applications of HCL.

To overcome this problem, four different prediction intervals (two frequentist, two Bayesian), 
that can be applied to such data, are proposed. 
Based on comprehensive Monte-Carlo simulations, the coverage probabilities of the 
proposed prediction intervals were compared to heuristical HCL typically used in 
daily toxicological routine (historical range, limits of the np-chart, mean $\pm$ 2 SD). 
Our simulations reveal, that frequentist bootstrap calibrated prediction intervals 
control the type-1-error best, but, also prediction intervals calculated based on Bayesian 
generalized linear mixed models appear to be practically applicable. Contrary, all
heuristics fail to control the type-1-error. 

The application of HCL is demonstrated based on a real life data set containing historical controls
from long-term carcinogenicity studies run on behalf of the U.S. National Toxicology Program. 
The proposed frequentist prediction intervals are publicly available from the R package predint,
whereas R code for the computation of the two Bayesian prediction intervals is provided via GitHub. \\

\textit{Keywords}: Bootstrap-calibration, micro-nucleus-test, long-term carcinogenicity studies,
OECD test guideline, Shewhart control chart, Bayesian hierarchical modeling

\newpage

\section{Introduction}

\subsection{Historical control data (HCD)}
For several toxicological and pre-clinical study types it is mandatory to store and 
report the outcome of the negative and sometimes also positive control groups of
the experiments conducted before the current study (see e.g. OECD guidelines 471, 487,
489 or EFSA 2024).  
In this context it is required to validate the concurrent control group by so called historical 
control limits (HCL) which, based on the HCD aim to predict the outcome of the current control group
with x \% (usually 95 \%) of confidence (Kluxen et al. 2021, Dertinger et al. 2023, Menssen 2023).

This approach 
is seen as a quality check for the current control group(s) and it is interpreted 
as a warning signal, if the current control is not covered by the HCL (Vandenberg et al. 2020):
Either as a warning for a 
false-positive result of the current study (if the concurrent control falls below the lower HCL)
or as a warning for a false-negative result (if the concurrent control falls above the upper HCL).
With other words, it is desired that the HCL cover the central 95 \% of the possible values that can be obtained from the true underlying distribution.

Despite the fact, that the HCL based validation of a current control is required by several test guidelines 
(e.g. OECD 471, OECD 473, OECD 487, OECD 490),
most guidelines fail to provide reproducible methodology on how to compute HCL. Hence, many heuristics
such as the historical range (the min and the max of the HCD) or the mean $\pm$ 2 standard deviations 
are used for the practical calculation of
HCL, but suffer from severe theoretical drawbacks (Menssen 2023). 

\subsection{Calculation of historical control limits based on prediction intervals}
Contrary to the widely applied heuristics, prediction intervals directly aim  
to predict one (or more) value(s) from the same data generating process from which
 the historical data is derived. Hence several authors recommend the use of prediction intervals for the
application of HCL (EFSA 2024, Menssen and Schaarschmidt 2019, Menssen and Schaarschmidt 2022,
Menssen 2023, Kluxen et al. 2021, Dertinger et al. 2023). 

It has to be noted that the prediction intervals presented in several standard textbooks and other 
scientific publications (e.g. Hahn and Meeker 1991, Hahn et al. 2017, Wang 2008, Tian et al. 2022, Nelson 1982) ground on the assumption that the historical sample is comprised of independent and identically distributed (iid) observations, meaning that possible between-study variation is assumed to be absent.
With other words, this simplistic prediction intervals do not account for the hierarchical 
design of the historical control data, i. e. they do not reflect that certain experimental units belong to a certain historical control group of a certain historical study. 

Hence, the restriction to the iid assumption does not make sense in the context of toxicological HCD for which systematic between study variation is a fundamental part of the data generating process, and consequently, is regularly reported in real life data. For example, HC data sets that are prone to systematic between study variation can be found in Tarone 1982, Carlus et al. 2013, Menssen et al. 2024, Levy et al. 2019, Tug et al. 2024, Dertinger et al. 2023 (for further details see section 4 of the supplementary material).

\subsection{Dichotomous endpoints}
In toxicology, several study types assess dichotomous endpoints: e.g. carcinogenicity studies (rats with vs.
rats without a tumor), the micro nucleus test (cells with vs. cells without a micronucleus)
or the recently developed liquid microplate fluctuation Ames MPF$^{TM}$ test 
following Spiliotopoulos et al. 2024 (numbers of wells with vs. number of wells without a color change).

It seems natural to model such data based on the binomial distribution. But, as described before, the
hierarchical design of the HCD gives rise for systematic between-study variation. Thus, the variability 
of the data will exceed the variance that can be modeled 
by the simple binomial distribution. This effect is called overdispersion (or extra-binomial variation) 
and can either be modeled based on the quasi-binomial assumption - which relies on the assumption that 
the between-study overdispersion is constant - or based on the beta-binomial distribution in which the between-study overdispersion depends on the number of experimental units within each HC group (see section 2). 

In the following, we provide four different methods to compute prediction intervals for overdispersed
binomial data: Two frequentist intervals that depend on normal approximation and bootstrap calibration as well as two Bayesian intervals of which one is based on Bayesian hierarchical modeling, whereas the other depends on the Bayesian version of a generalized linear random effects model.

The manuscript is organized as follows: 
The next section provides an overview about the modeling of overdispersed binomial endpoints.
Section \ref{sec::control_limits} provides an overview about the calculation of HCL.
Several heuristical HCL that are used in daily toxicological routine are given in section \ref{sec::heristics}. The proposed prediction intervals are described in sections \ref{sec::bs_pi}
and \ref{sec::bayes_pi}. 
Section \ref{sec::real_life_data} reviews the properties of two real life HC data bases,
that serve as an inspiration for Monte-Carlo simulations regarding the coverage probabilities
of the proposed methodology (section \ref{sec:simulation}). The application of
the proposed methods is demonstrated in section \ref{sec::application}. The manuscript ends
with a discussion (section \ref{sec:discussion}) and conclusions (section 
\ref{sec::conclusions}).


\section{Models for overdispersed binomial data} \label{sec:Model}

One possible way to model overdispersion is the beta-binomial distribution
in which the binomial proportion for each of the $h=1, \ldots , H$ historical control groups 
is derived from a beta distribution
\begin{equation}
	\pi_h \sim Beta(a, b) \label{eq::pi_beta}
\end{equation}
and the random variable within each cluster is binomial
\begin{equation}
	Y_h \sim Bin(n_h, \pi_h). \label{eq::y_beta_bin}
\end{equation}
In this setting $E(\pi_h)= \pi = a/(a+b)$, $E(Y_h)=n_h \pi$. Furthermore, $\pi$ is the overall 
binomial proportion and
$n_h$ is the a priori known and fixed total number of experimental units per historical control group. 
The variance of the beta-binomial distributed random variable is
\begin{equation}
	var(Y_h)= n_h \pi(1-\pi) \big[1 + (n_h - 1)\rho \big]. \label{eq::varBB}
\end{equation}
In this parametrisation $\rho=1/(1+a+b)$ describes the intra-class correlation coefficient
(Lui et al. 2000).
Because the parameters of the beta distribution are restricted to be $a>0$ and 
$b>0$, also the intra-class correlation is restricted to be $0 < \rho < 1$. \\
Another possibility to model between-study overdispersion is the quasi-binomial
approach in which the dispersion parameter $\phi$ constantly inflates the variance
\begin{equation}
	var(Y_h)= n_h \pi(1-\pi) \phi. \label{eq::varQB}
\end{equation}
In the case in which the number of experimental unis within each control
group becomes equal $n_h = n_{1}$, also the part of the beta-binomial variance 
that controls the magnitude of overdispersion per control group 
($\big[1 + (n_h - 1)\rho \big]$ in eq. \ref{eq::varBB}) becomes a constant. 
In this condition, the beta-binomial model can be interpreted as a special case of the 
quasi-binomial model - one in which the magnitude of the intra-class correlation $\rho$ 
together with the cluster size $n_h=n_1$ determines the magnitude of the dispersion parameter $\phi$. 
Consequently, in this special case, both models coincide (Menssen 
and Schaarschmidt 2019) and
\begin{equation}
	\phi \widehat{=} \big[1 + (n_h - 1)\rho \big] \text{ if all } n_h = n_{1}.\label{eq::overdisp}
\end{equation}
Both approaches are standard for modeling overdispersion (McCullagh and Nelder 1989),
but in practice it might sometimes appear by chance, that a data set shows signs of underdispersion 
(less variability than possible under the simple binomial assumption). 
From eq. \ref{eq::varBB} it can be deduced that underdispersion is caused by negative correlations
between the experimental units within each cluster (if the intra-class correlation $\rho$
was allowed to be negative). Practically this would mean that the chance to observe a 
further experimental unit with an event decreases, with the occurrence of an event (e.g.
if one rat develops a tumor, the chance that the other rats in the cohort develop a tumor decreases). 
From a biological point of view, this is heavily implausible. Consequently the quasi-binomial 
dispersion parameter was restricted to be $\phi \geq 1$ in the following sections of this manuscript.

\section{Historical control limits for overdispersed binomial data} \label{sec::control_limits}
The calculation of the historical control limits $[l, u]$ given below is based on
the observed number of experimental units with an event $y_h$ out of a total number 
of experimental units per historical control group $n_h$ (e.g. rats with a tumor out of a total number of rats). 
All historical control limits (HCL) are aimed to cover a further number of experimental 
units with an observed number of successes $y^*=y_{H+1}$ out of a further a priori known total 
number of experimental units $n^*=n_{H+1}$ with coverage probability 
\begin{equation}
	P(l \leq y^* \leq u) = 1-\alpha.
\end{equation}
Since HCL are computed to validate the current control group, the desired HCL should aim to cover the central $100(1-\alpha)$\% of the underlying distribution and its limits should approach the lower $\alpha/2$ and the upper $1-\alpha/2$ quantile, respectively (see fig. 1 of Francq et al. 2019). It it crucial to ensure for this behavior also, if the underlying distribution is skewed (Hoffman 2003). Consequently, the desired interval has to ensure for equal tail probabilities
\begin{equation}
	P(l \leq y^*) = P( y^* \leq u) = 1-\alpha / 2.
\end{equation}
This means that in practice, the control limits should cut off the lower and 
the upper $100(\alpha/2)$\% of the underlying distribution, regardless 
if this distribution is symmetrical or skewed. For overdispersed
binomial data, possible skewness depends on three factors: 
\begin{enumerate}
	\item The skewness increases, the closer $\pi$ approaches its boundaries (0 or 1). 
	\item The skewness tends to increase with increasing between-study overdispersion 
	(depending on the value for $\pi$). 
	\item The skewness also tends to increases with decreasing cluster size $n^*$ 
	(depending on the values for $\pi$ and the amount of between-study overdispersion). 
\end{enumerate}

This behavior is demonstrated in fig. 1 and 2 of the supplementary material.


\subsection{Heuristical historical control limits} \label{sec::heristics}

Although several authors, dissuade from its use (Greim et al. 2003, 
Kluxen et al. 2021, EFSA 2024, Menssen 2023, Menssen et al. 2024), HCL based on the historical
range are frequently applied in daily toxicological routine (Zarn et al. 2024, 
WHO 2023, NTP 2018, EFSA 2010). Since such HCL are given by
\begin{equation}
	\Big[ l=min(y_h), u=max(y_h) \Big]
\end{equation}
they aim to cover \textit{all} realizations from the underlying distribution and
hence, do not aim for a proper approximation of its central $100(1-\alpha)$\%. 
Note that the min and max of the observed data provide very poor estimates for 
the true min and max of the underlying distribution, especially if the number of historical 
observations is low. This is because values in the center of the distribution have a higher 
chance to occur, whereas the occurrence of more extreme values in the tails of the distribution 
are less likely.
Furthermore it has to be noted, that this interval is explicitly based on the assumption
of constant cluster size $n_h=n^* \text{ } \forall \text{ } h=1, \ldots H$, because
the interpretation of the number of observed events makes only sense in relation to the given cluster size
(meaning that 5 out of 50 rats with a tumor have a fundamental different interpretation
than 5 out of 100 rats). Hence, a sensible interpretation of the historical range is
not possible, if the cluster size differs between the studies.

Another popular method for the calculation of control limits for dichotomous data 
are the limits used in Shewhart np-charts
\begin{equation}
	[l,u]= n^* \bar{\pi} \pm k \sqrt{n^* \bar{\pi} (1-\bar{\pi})}
\end{equation}
with $\bar{\pi} = \frac{\sum_h y_h}{\sum_h n_h}$. In practical applications $k$ is usually
set to 2 or 3 in order to approximate the central 95.4 \% or 99.7 \% of the underlying 
distribution (Montgomery 2019). This HCL are strictly based on the assumption 
that the observations are independent realizations of the same binomial distribution
which can be adequately approximated by a normal distribution with mean $n^* \bar{\pi}$
and variance $n^* \bar{\pi} (1-\bar{\pi})$. Hence, the Shewhart np-chart presumes that 
the law of large numbers is applicable, ignores the variability in $\bar\pi$, 
fails to account for overdispersion and fails to ensure for equal tail
probabilities. 

HCL that are based on the mean $\pm$ $k$ standard deviations are given by
\begin{equation}
	[l,u]= \bar{y} \pm k SD
\end{equation}
with $\bar{y}=\frac{\sum_h y_h}{H}$ and $SD= \sqrt{\frac{\sum_h (\bar{y}-y_h)^2}{H-1}}$. 
This type of HCL is based on a simple normal approximation that lacks an explicit assumption 
about the mean-variance relationship and hence heuristically allows for overdispersion (Menssen et 
al. 2024). Similar to HCL that are based on the historical range, the mean $\pm$ $k$ standard 
deviations are explicitly based on the assumption of constant cluster size 
$n_h = n^* \quad \forall  \quad h= 1, \ldots , H$,
because for dichotomous endpoints, the number of observed experimental units with an event is only interpretable 
in the context of the total number of experimental units in the particular control group $n_h$.
Hence, the application of HCL that are computed by the mean $\pm$ $k$ standard deviations
to dichotomous HCD with different cluster sizes should strictly be avoided
(see section 2 of the supplementary material).

\subsection{Bootstrap calibrated prediction intervals} \label{sec::bs_pi}

Similarly to the prediction intervals for overdispersed count data of Menssen et al. 2024,
the two frequentist prediction intervals below are based on the assumption that
\begin{equation*}
	\frac{n^* \hat{\pi}- Y^*}{\sqrt{var(n^* \hat{\pi}) + var(Y^*)}} \stackrel{\text{approx.}}{\sim} N(0, 1).
\end{equation*}
Here, $Y^*$ is the future random variable (of which $y^*$ is a realization),
 $\hat{\pi}$ is the binomial proportion estimated
from the HCD and $n^*$ is the a priori known number of experimental units of the future control 
group and
$n^* \hat{\pi}$ and $Y^*$ are assumed to be independent random variables. 

Following Menssen and Schaarschmidt 2019, a prediction interval which is based on the quasi-binomial assumption is given by
\begin{equation}
        [l,u]=n^* \hat{\pi} \pm z_{1-\alpha/2} \sqrt{\frac{\hat{\phi} n^{*2} \hat{\pi} (1- \hat{\pi})}{\sum_h n_h} +
        \hat{\phi} n^* \hat{\pi} (1- \hat{\pi})} \label{eq:simple_pi_QB}
\end{equation} 
with $\hat{\phi}>1$ as an estimate for the between-study overdispersion.

If between-study overdispersion is modeled based on the beta-binomial 
distribution, the corresponding prediction interval is given by 
\begin{gather}
        [l,u]=n^* \hat{\pi} \pm z_{1-\alpha/2} \sqrt{\Big[\frac{n^{*2} \hat{\pi} (1- \hat{\pi})}{\sum_h n_h} + 
        \frac{\sum_h n_h -1}{\sum_h n_h} n^{*2} \hat{\pi} (1- \hat{\pi}) \hat{\rho}\Big]+
        n^* \hat{\pi} (1- \hat{\pi}) [1 + (n^* -1) \hat{\rho}]} \label{eq:simple_pi_BB}
\end{gather} 
with $\hat{\rho}$ as an estimate for the intra-class correlation coefficient.
Details on methodology for the estimation of $\hat{\pi}$, $\hat{\phi}$ and $\hat{\rho}$ 
are given in section \ref{sec::comp_det}.

As mentioned above, overdispersed binomial data can be heavily right- or left-skewed, 
but the prediction intervals in eq. \ref{eq:simple_pi_QB} and \ref{eq:simple_pi_BB} are
still symmetrical. Furthermore, they depend on normal approximation and hence, 
tend to have coverage probabilities below the nominal level, especially if the 
number of historical studies is low.

To overcome this problem, a bootstrap calibration algorithm (see box below), 
that individually calibrates both interval borders was applied. 
Note, that the idea behind this algorithm is similar to a Monte-Carlo simulation
regarding the prediction intervals coverage probability. 
Therefore, the bootstrap samples $\boldsymbol{y}_b$ that mimic the HCD, as well the 
bootstrapped future observation $y^*_b$ are derived from the same data generating process
(which depends on the estimates for the model-parameters estimated from the HCD). 
The algorithm aims to provide substitutes $q_l$ and $q_u$ for the standard normal 
quantile used in eq. \ref{eq:simple_pi_QB} and \ref{eq:simple_pi_BB},
such that possible skewness of the underling distribution is taken into account to ensure 
for equal tail probabilities and enhance the intervals coverage probability.

\begin{framed}
\textbf{Bootstrap calibration of the proposed prediction intervals}
\begin{enumerate}
        \item Based on the historical data $\boldsymbol{y}=(y_1, \ldots, y_H)$ find
        estimates for the model parameters ($\hat{\pi}$ and $\hat{\phi}$ in the quasi-binomial 
        case and $\hat{\pi}$ and $\hat{\rho}$ in the beta-binomial case
        \item Based on these estimates, sample $B$ parametric bootstrap 
        samples $\boldsymbol{y}_b$ following the same experimental design as the 
        historical data (for sampling algorithms see section 3 of the 
        supplementary material)
        \item Draw $B$ further bootstrap samples $y^*_b$ that mimic the possible numbers of
        success in a future control group of size $n^*$ (using the same sampling algorithms as in 
        step 2)
        \item Based on $\boldsymbol{y}_b$ obtain bootstrapped estimates for the model parameters 
        (either $\hat{\pi}_b$ and $\hat{\phi}_b$ or $\hat{\pi}_b$ and $\hat{\rho}_b$) and calculate 
        $\widehat{var}_b(n^* \hat{\pi})$ and $\widehat{var}_b(Y^*)$
        \item Calculate lower prediction borders $l_b = n^* \hat{\pi}_b - q_l 
        \sqrt{\widehat{var}_b(n^* \hat{\pi}) + \widehat{var}_b(Y^*)}$, such that
        all $l_b$ depend on the \textit{same} value for $q_l $.
        \item Calculate the bootstrapped coverage probability 
        $\hat{\psi}_l=\frac{\sum_b}{B} I_b \text{ with } 
        I_b = 1 \text{ if } l_b \leq y^*_b \text{ and }
        I_b = 0 \text{ if } y^*_b < l_b$
        \item Alternate $q_l$ until $\hat{\psi}_l \in (1-\frac{\alpha}{2}) \pm t$ 
        with $t$ as a predefined tolerance around $1-\frac{\alpha}{2}$
        \item Repeat steps 5-7 for the upper prediction border with 
        $\hat{\psi}_u=\frac{\sum_b}{B} I_b \text{ with } 
        I_b = 1 \text{ if } y^*_b \leq u_b \text{ and }
        I_b = 0 \text{ if } u_b < y^*_b$
        \item Use the corresponding coefficients $q^{calib}_l$ and $q^{calib}_u$ and the estimates
        obtained from the initial HCD (either $\hat{\pi}$ and $\hat{\phi}$
        or $\hat{\pi}$ and $\hat{\rho}$) for interval calculation
        \begin{gather*}
                \big[l= n^* \hat{\pi} - q^{calib}_l 
                \sqrt{\widehat{var}(n^* \hat{\pi}) + \widehat{var}(Y^*)},\\
                \quad\quad\quad\quad\quad 
                u= n^* \hat{\pi} + q^{calib}_u
                \sqrt{\widehat{var}(n^* \hat{\pi}) + \widehat{var}(Y^*)} \big]
        \end{gather*}
\end{enumerate}
\end{framed}
The bootstrap calibrated quasi-binomial prediction interval is given by
\begin{gather}
\begin{aligned}
        \Big[l=n^* \hat{\pi} - q^{calib}_l \sqrt{\frac{\hat{\phi} n^{*2} \hat{\pi} (1- \hat{\pi})}{\sum_h n_h} +
        \hat{\phi} n^* \hat{\pi} (1- \hat{\pi})} \\ 
        \quad \quad \quad
         u=n^* \hat{\pi} + q^{calib}_u \sqrt{\frac{\hat{\phi} n^{*2} \hat{\pi} (1- \hat{\pi})}{\sum_h n_h} +
        \hat{\phi} n^* \hat{\pi} (1- \hat{\pi})} \Big]. 
\label{eq:quasi_bin_pi_calib}
\end{aligned}
\end{gather} 
Similarly, the bootstrap calibration of the beta-binomial prediction interval results in
\begin{gather}
\begin{aligned}
        \Big[l=n^* \hat{\pi} - q^{calib}_l \sqrt{n^* \hat{\pi} (1- \hat{\pi}) [1 + (n^* -1) \hat{\rho}] +
        [\frac{n^{*2} \hat{\pi} (1- \hat{\pi})}{\sum_h n_h} + 
        \frac{\sum_h n_h -1}{\sum_h n_h} n^{*2} \hat{\pi} (1- \hat{\pi}) \hat{\rho}]} \\
        \quad \quad \quad
        u=n^* \hat{\pi} + q^{calib}_u \sqrt{n^* \hat{\pi} (1- \hat{\pi}) [1 + (n^* -1) \hat{\rho}] +
        [\frac{n^{*2} \hat{\pi} (1- \hat{\pi})}{\sum_h n_h} + 
        \frac{\sum_h n_h -1}{\sum_h n_h} n^{*2} \hat{\pi} (1- \hat{\pi}) \hat{\rho}]} \Big].
\label{eq:beta_bin_pi_calib} 
\end{aligned}
\end{gather} 
The applied bootstrap calibration is a modified version of the algorithm 
proposed by Menssen and Schaarschmdit 2022 and has also been applied to  
calibrate prediction intervals for 
overdispersed count data (Menssen et al. 2024).
The only difference to Menssen and Schaarschmdit 2022 is, that both interval 
limits are calibrated individually, but the search for $q^{calib}_l$ and $q^{calib}_u$
depends on the same bisection procedure using a tolerance $t=0.001$.\\

\subsection{Bayesian modeling} \label{sec::bayes_pi}

Bayesian approaches provide an alternative estimation of prediction intervals (Hamada et al. 2004).
 Future observations as well as all model parameters are interpreted as random variables depending on each other. Thus, a posterior distribution of a future observation can be derived immediately within the parameter estimation process. 

\subsubsection{Hierarchical modeling}

Bayesian hierarchical modeling depends on the beta-binomial model that is given in equations \ref{eq::pi_beta} 
and \ref{eq::y_beta_bin}.
In a Bayesian interpretation, the experiments' parameters $\pi_h$ are assumed to follow a Beta distribution, as a second level in a hierarchy (Howley and Gibbert 2003). 

A Beta prior in mean-precision-parametrization (Beta proportion distribution, Stan development Team 2024a) is applied: 
$$\pi_h \sim \text{Beta}(\mu, \kappa) \quad \forall\; h=1, \ldots, H$$
with a non-informative prior for the location hyperparameter $\mu \in (0,1)$. A weakly informative gamma prior $\kappa \sim \text{Ga}(a,\: b)$ should be applied to represent a realistic domain of the precision hyperparameter $\kappa>0$ and, thus, to obtain a stable estimation.  

The parameters' posterior distributions, given the data, i.e. symbolically   
\begin{gather*}
p(\pi_1, ..., \pi_H, \mu, \kappa \: | \: Y_1, ..., Y_H)  \propto p(Y_1, ..., Y_H \: | \: \pi_1, ..., \pi_H) \: p(\pi_1, ..., \pi_H \: | \: \mu, \kappa) \: p(\mu, \kappa)=\\
p(\mu, \kappa) \prod_{h=1}^H p(\pi_h \: | \: \mu, \kappa) \: p(Y_h \: | \: \pi_h)
\end{gather*}
are estimated by Markov-chain-Monte-Carlo (MCMC) sampling. 

A posterior predictive distribution of a future observation $y^*$ is obtained by random sampling per MCMC iteration $c=1, ..., C$: For every MCMC sample $(\tilde\mu_c, \tilde\kappa_c)$ of $\mu$ and $\kappa$, a prediction $$\tilde\pi^*_c \sim \text{Beta}(\tilde\mu_c, \tilde\kappa_c)$$ of the future experiment's success proportion is drawn, and in turn a predicted future observation is sampled as $$\tilde{y}^*_c \sim \text{Bin}(n^*, \tilde\pi^*_c).$$ A pointwise prediction interval for one future observation $y^*$ is then obtained using empirical quantiles of $\left\{\tilde{y}^*_1, ..., \tilde{y}^*_C \right\}$ such that $$[l, u]= [q_{\alpha/2}(\left\{\tilde{y}^*_1, ..., \tilde{y}^*_C \right\}),  q_{1-\alpha/2}(\left\{\tilde{y}^*_1, ..., \tilde{y}^*_C \right\})]$$ in the two-sided case.

\subsubsection{Bayesian generalized linear mixed model}

A concept closely related to the Beta-binomial model shown above is a generalized linear mixed model (GLMM), where Bernoulli distributed random variables $Z_{h1}, ..., Z_{hn_h}$ with $\sum_{k=1}^{n_h} Z_{hk} = Y_h$ are considered (Fong et al. 2010). The success proportion $\pi_h = E(Z_{hk})$ is linked to a linear predictor $$\eta_h = \ln \frac{\pi_h}{1 - \pi_h} \in (-\infty, +\infty ),$$ 
which is comprised of a fixed general intercept $\nu$ and random effects $\beta_h$ per experiment: 
$$\eta_h = \nu + \beta_h, \quad \beta_h \sim N(0, \sigma).$$ 

With that, the historical experiments are again assumed to have individual success proportions $\pi_h$, derived from the same distribution, and the observations are again realizations of the binomial random variable $Y_h \sim \text{Bin}(n, \pi_h)$.

The posterior distributions of $\nu \in (-\infty, +\infty )$ and $\sigma > 0$ are estimated by MCMC sampling using non-informative vague priors. Realizations of the future random variable $Y^* = \sum_{k=1}^{n^*} Z^*_k$ are derived:
For every MCMC sample $c$, a prediction 
$$\tilde\beta^*_c \sim N(0, \tilde\sigma_c)$$ 
of the future experiment's random intercept is drawn, followed by
$$\tilde\eta^*_c = \tilde\nu_c + \tilde\beta^*_c.$$ 
Then, the future experiment's success proportion 
$$\tilde\pi^*_c = \frac{\exp(\tilde\eta^*_c)}{\exp(\tilde\eta^*_c) + 1}$$
is derived, and in turn $$\tilde{y}^*_c \sim \text{Bin}(n^*, \tilde\pi^*_c)$$ is drawn from the latter. 
Similarly as above, a pointwise two-sided prediction interval for one future observation $y^*$ 
is given by $$[l, u]= [q_{\alpha/2}(\left\{\tilde{y}^*_1, ..., \tilde{y}^*_C \right\}),  q_{1-\alpha/2}(\left\{\tilde{y}^*_1, ..., \tilde{y}^*_C \right\})].$$ 

\subsection{Computational details and estimation} \label{sec::comp_det}

The proposed frequentist prediction intervals are implemented in the R package 
\texttt{predint} (Menssen 2023). The uncalibrated prediction intervals (eq. \ref{eq:simple_pi_QB} and \ref{eq:simple_pi_BB}) are provided via \texttt{qb\_pi()}
and \texttt{bb\_pi()}. The bootstrap calibrated prediction intervals (eq. \ref{eq:quasi_bin_pi_calib}  and \ref{eq:beta_bin_pi_calib} )
are implemented via the functions \texttt{quasi\_bin\_pi()} and \texttt{beta\_bin\_pi()}.

The estimates $\hat{\pi}$ and $\hat{\phi}$ used for the quasi-binomial prediction
interval were estimated based on a generalized linear model (\texttt{stats::glm()}).
The estimates $\hat{\pi}$ and $\hat{\rho}$ for the beta-binomial distribution 
were obtained following Lui et al. 2000. In order to avoid undersipersion, the 
quasi-binomial interval was computed with $\hat{\phi}\geq 1.001$. The estimated 
intra-class correlation used for the beta-binomial interval was restricted to 
$\hat{\rho} \geq 0.00001$.

The bisection procedure used for bootstrap calibration is implemented in \texttt{predint} and provided via the function \texttt{bisection()}. This function takes three different lists as input, that contain the 
bootstrapped expected observations $\hat{y}^*_b$, the bootstrapped standard errors
$\sqrt{\widehat{var}({n^* \hat{\pi})_b + \widehat{var}(Y^*)_b}}$ and the bootstrapped further observations $y^*_b$ and returns values 
for $q_l$ and $q_u$. Hence, this function enables the calculation of bootstrap 
calibrated prediction intervals in a general way, such that any Wald-type prediction
interval can be calibrated, as long as a parametric model can be fit to the data 
and a sampling function is available based on which parametric bootstrap samples 
can be generated.

The Bayesian hierarchical beta-binomial model was applied using the software 
\texttt{stan} via the R package \texttt{rstan} (Stan Development Team 2024). 
In the simulation, a weakly informative prior $\kappa \sim \text{Ga}(2,\: 5\cdot 10^{-5})$
was used, giving the domain of the precision hyperparameter $\kappa>0$.
The Bayesian GLMM was applied using the R function \texttt{stan\_glmer} 
from the package \texttt{rstanarm} (Goodrich et al. 2024). Functions for 
the computation of both Bayesian prediction intervals are provided via GitHub 
\url{https://github.com/MaxMenssen/pi_overdisp_binomial/tree/main}.


\section{Properties of real life HCD} \label{sec::real_life_data}

This section reviews the statistical properties of two real life HC data bases:
One about long-term carcinogenicity studies and one about the micronucleus test.
The properties of the two data bases were taken into account tor the setup of the 
Monte-Carlo simulations regarding the coverage probability of the different HCL.
This two study types were chosen in order to reflect the different properties of 
overdispersed binomial data at very different locations in the possible parameter 
space in order to enhance the simulations generalizability.


\subsection{HCD from the micro-nucleus test} 

A micronucleus is a small fragment of chromosomes remaining outside the new nuclei 
after cell division. In the in-vitro micro-nucleus test (MNT), cell cultures on a 
well plate are treated with the compound of interest and an increased number of cells 
with a micronucleus is used as a proxy for genotoxicity (Fenech 2000). Negative and 
positive control wells are observed to measure the spontaneous micronucleus proportion 
without treatment and to evaluate the test assay’s functionality by comparison to HCD.

The analyzed real life HCD includes negative and positive control 
groups from about $H \approx 50$ experiments. Due to the high extend of
standardization, the aggregated number of scored cells per control group is
always $n_h=n^*=18000$.

The HC data base hints at an average proportion of cells with at least one micronucleus 
of about $\hat{\pi} \approx 0.01$ for negative control groups and about $\hat{\pi} \approx 0.15$ to $0.2$ 
for positive controls. Overdispersion was estimated to range between $\hat{\phi} \approx 10$ 
and $50$ for negative controls and tends to reach several hundreds for positive controls. 

 
\subsection{HCD from long-term carcinogenicity studies}

Historical control data on different endpoints obtained in long-term carcinogenicity 
studies (LTC) are publicly available via the Historical Control Database of the U.S. 
National Toxicology Program (NTP 2024). From this resource Menssen and Schaarschmidt 2019 
extracted HCD about the numbers of B6C3F1 mice per control group that had a hemangioma, 
the numbers of mice that showed at least one malignant tumor as well as the numbers of 
mice that passed away before the end of the study. This data was analyzed 
with regard to the binomial proportion $\hat{\pi}$ and the amount of overdispersion 
$\hat{\phi}$.

For rare events (hemangioma) the reported proportion was relatively low (close to zero) 
and overdispersion
was estimated to be absent in most of the cases. The proportion of animals 
with malignant tumors ranged between 0.2 and 0.8, whereas the estimated dispersion
parameters raised up to roughly four, meaning that the particular data set is 
four times as variable than possible under simple binomial distribution.
The reported mortality rates behaved in a similar fashion. All reported control groups
were comprised of 50 animals.


\section{Simulation study} \label{sec:simulation}

In order to asses the coverage probabilities of the different methods for the 
calculation of HCL reviewed above, Monte-Carlo simulations were run. In all 
simulations, the nominal coverage probability was  set to $1-\alpha=0.95$. 
The parameter combinations used for the simulation were inspired by the real 
life data shown above. For all simulations the cluster size was fixed 
($n_h=n^* \quad \forall \quad h=1, \ldots , H$) such that the variance-mean 
relationships of the beta-binomial distribution and the quasi-binomial assumption 
coincide. \\
For each combination of model parameters, $S=5000$ "historical" data sets were drawn,
based on which historical control limits $[l, u]_s$ were calculated. 
Furthermore, $S$ sets of single "future" observations $y^*_s$ were sampled and the
coverage probability of the control intervals was computed to be 
\begin{gather}
        \hat{\psi}^{cp}=\frac{\sum_{s=1}^S I_s}{S} \text{ with}  \label{eq::psi_cp} \\
        I_s = 1 \text{ if } y^*_s \in [l, u]_s, \notag \\
        I_s = 0 \text{ if } y^*_s \notin [l, u]_s. \notag
\end{gather}
Average interval borders were calculated as the mean of the simulated interval borders 
$\bar{l}=\frac{\sum_{s=1}^S l_s}{S}$ and $\bar{u}=\frac{\sum_{s=1}^S u_s}{S}$.
Each single bootstrap calibrated prediction interval that was calculated in the simulation
was based on $B=10000$ bootstrap samples. Each Bayesian prediction interval was calculated based on $C=5000$ MCMC samples.


\subsection{Coverage probabilities for the MNT-inspired simulation setting}

This part of the simulation was inspired by the properties of the real live HCD 
for the micro-nucleus test reviewed above. 72 combinations of values for the
number of historical studies $H$, the binomial proportion $\pi$ and the 
amount of overdispersion $\phi$ were run, of which 18 combinations directly 
reflect the properties of the real life HCD (combinations of the bold faced 
parameter values in tab. \ref{tab::mnt_setting}), whereas the remaining 54 
combinations of parameter values were run in order to enhance the generalisability
of the simulation.

Note that, because the cluster size was relatively huge ($n_h=n^*=18000$) most parameter combinations 
result in a relatively symmetrical underlying distribution (except for small proportions 
$\pi=0.001$ combined with high overdispersion $\phi=500$,
see supplementary fig. 1).

\begin{figure}[H]
  \includegraphics[width=\linewidth]{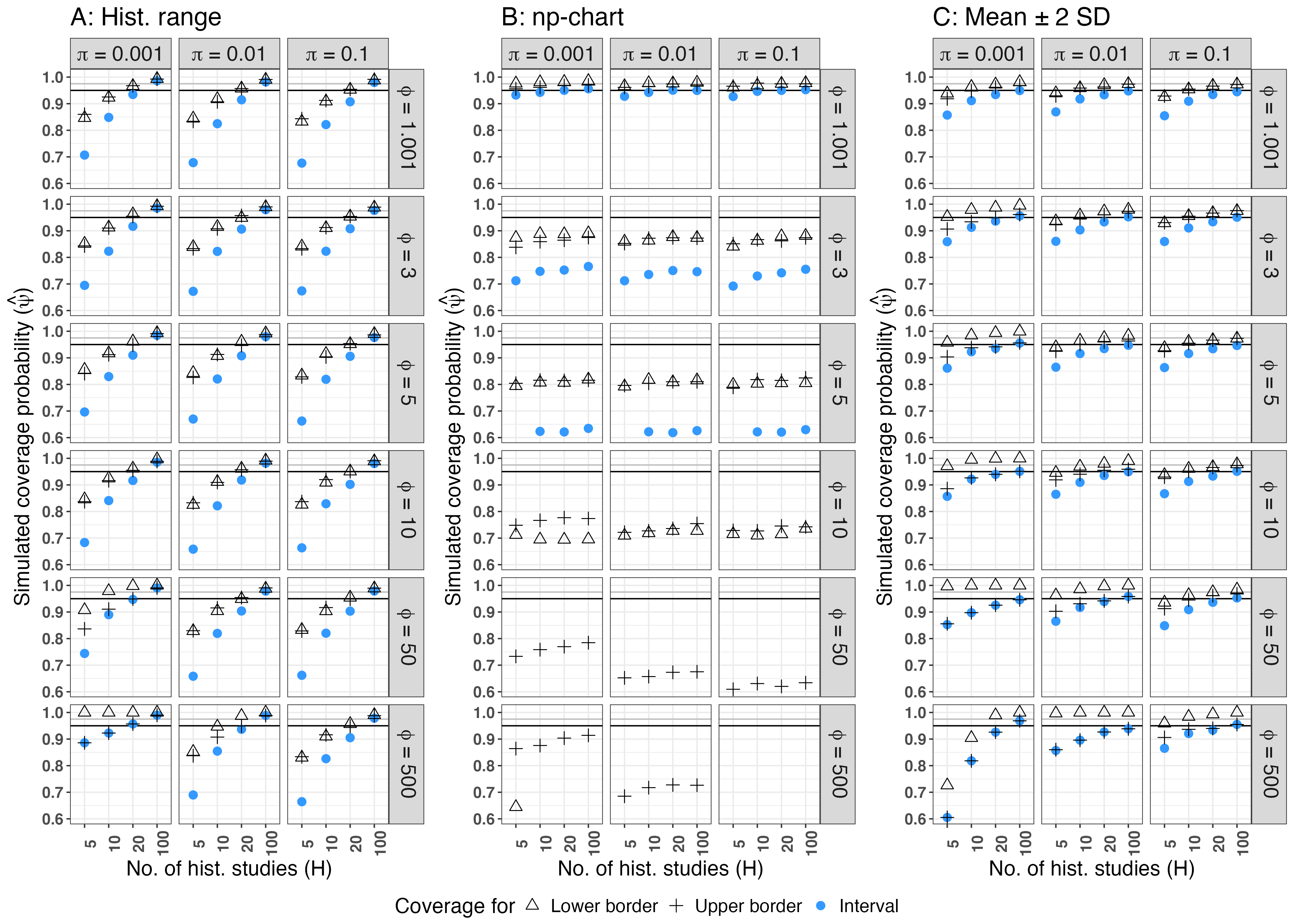}
  \caption{Simulated coverage probabilities of heuristical methods (MNT-setting).
  \textbf{A:} Historical range,
  \textbf{B:} np-chart,
  \textbf{C:} Mean $\pm$ 2 SD,
  \textbf{Black horizontal line:} Nominal coverage probability $\psi^{cp}=0.95$,
  \textbf{Grey horizontal line:} Nominal coverage probability for the lower and 
  the upper limit, if equal tail probabilities are achieved $\psi^{l}=\psi^{u}=0.975$.
  Coverage probabilities below 0.6 are excluded from the graphic.}
  \label{fig:heuristics_mnt}
\end{figure}

\begin{sidewaysfigure}
\centering
 	\includegraphics[width=\textwidth]{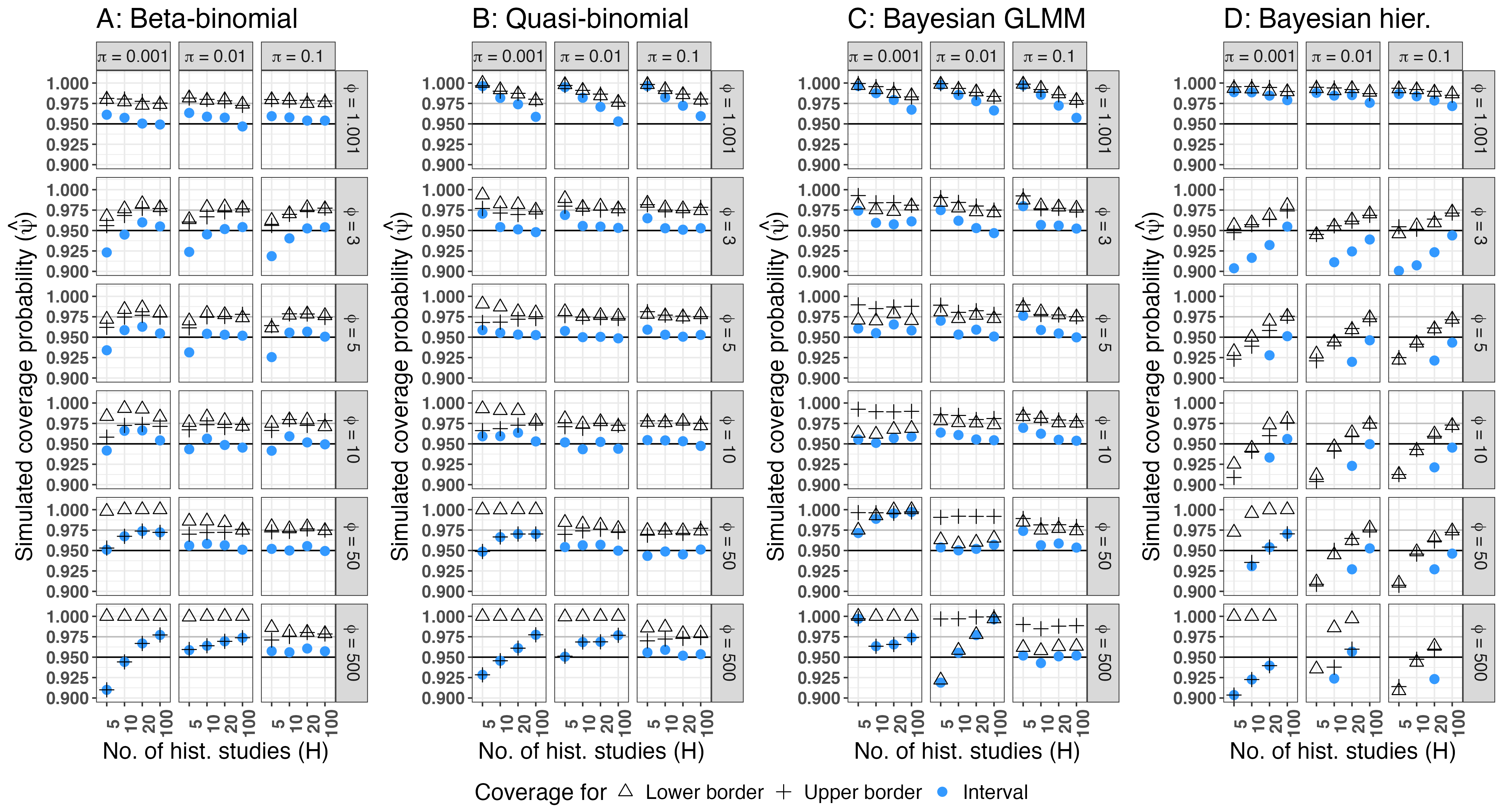}
  \caption{Simulated coverage probabilities of prediction intervals (MNT-setting).
  \textbf{A:} Calibrated beta-binomial prediction interval,
  \textbf{B:} Calibrated quasi-binomial prediction interval,
  \textbf{C:} Prediction interval obtained from a Bayesian generalized linear mixed model,
  \textbf{D:} Prediction interval obtained from a Bayesian hierarchical model,
  \textbf{Black horizontal line:} Nominal coverage probability $\psi^{cp}=0.95$,
  \textbf{Grey horizontal line:} Nominal coverage probability for the lower and 
  the upper limit, if equal tail probabilities are achieved $\psi^{l}=\psi^{u}=0.975$.
  Coverage probabilities below 0.9 are not shown. The Bayesian hierarchical model 
  faced serious convergence problems for $\phi=500$ and $H=100$. Hence, coverage probabilities 
  could not be computed in this setting.}
  \label{fig:predint_mnt}
\end{sidewaysfigure}

\begin{sidewaysfigure}
\centering
  \includegraphics[width=\linewidth]{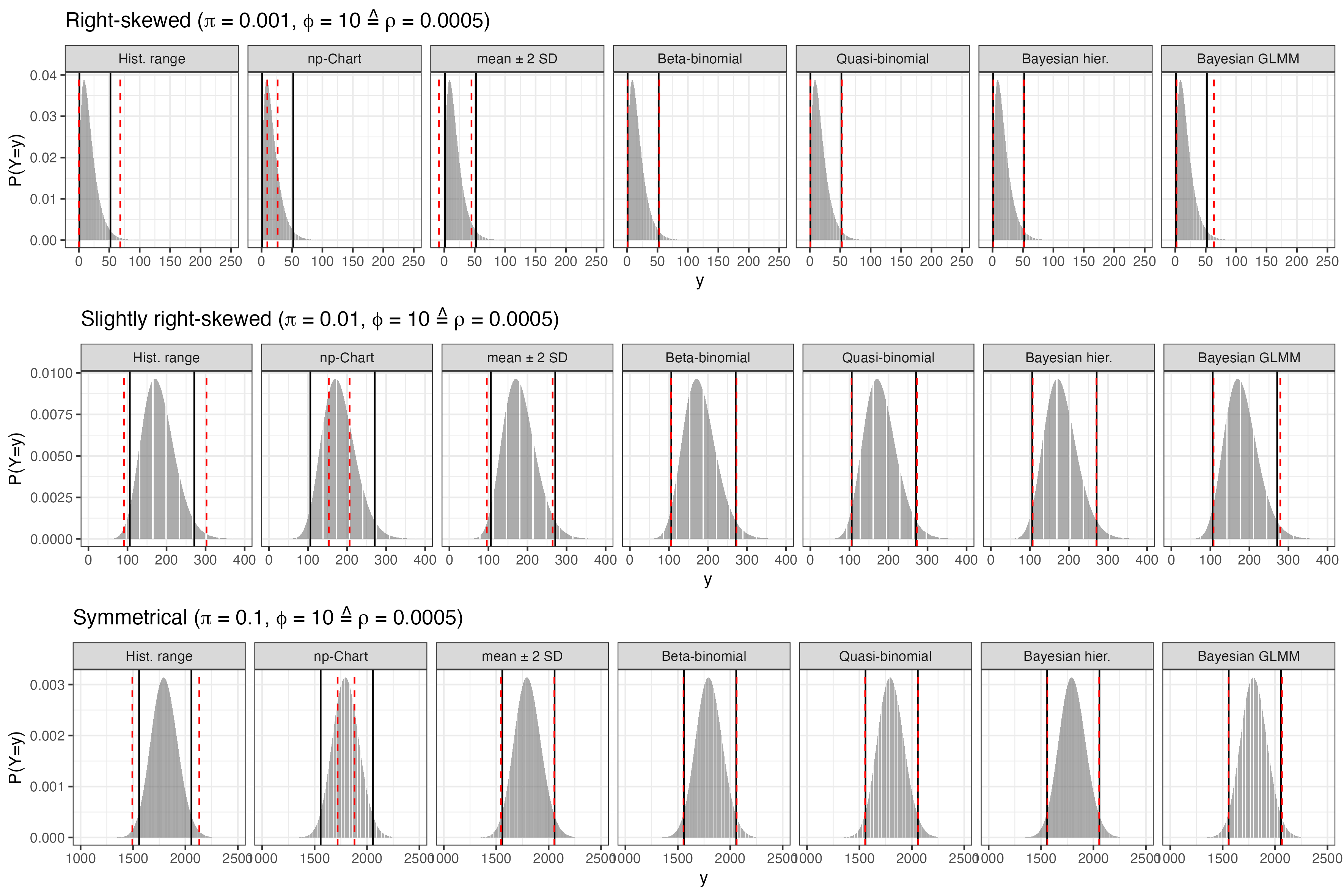}
  \caption{Average limits vs. true quantiles of the underlying distribution (MNT-setting).
  \textbf{Grey area:} Probability mass function of the underlying beta binomial distribution. 
  \textbf{Black lines:} True underlying 2.5 \% and 97.5\% quantiles.
  \textbf{Dashed red lines:} Average limits for $H=100$ and $n_h=18000$ obtained from the simulation.}
  \label{fig:hcl_vs_quantile_mnt}
\end{sidewaysfigure}

\begin{table}[h!]
\centering
\caption{Parameter values used for the MNT-inspired simulation} 
\begin{tabular}{|c|c|}
\hline 
\textbf{Parameter} & \textbf{Values}$^1$ \\ 
\hline 
No. of historical studies ($H$) & \textbf{5},  \textbf{10},  \textbf{20},  100 \\ 
\hline 
Binomial proportion ($\pi$) & 0.001,  \textbf{0.01},  \textbf{0.1} \\ 
\hline 
Overdispersion ($\phi$)$^2$ & 1.001, 3, 5,  \textbf{10},  \textbf{50},  \textbf{500} \\ 
\hline 
Historical cluster size ($n_h$) & \textbf{18000} \\ 
\hline 
Future cluster size ($n^*$) & \textbf{18000} \\ 
\hline 
\end{tabular} \\ [0.5ex]
\raggedright
\scriptsize
\textbf{Bold numbers:} Parameter values that reflect estimates obtained from real life HCD.
\textbf{1:} Simulations were run for all 72 combinations of the given parameter values.
\textbf{2:} Since the cluster sizes $n_h$ and $n°*$ are constant
$\phi \widehat{=} 1 + (n_h -1) \rho \widehat{=} 1 + (n^* -1) \rho$.
The given values of $\phi$ correspond to intra-class correlation coefficients $\rho$ of 
5.5e-08, 0.00011, 0.00022, 0.00050, 0.00272, 0.02772. 
\label{tab::mnt_setting}
\end{table}
The simulated coverage probabilities of the heuristical methods are depicted in fig. 
\ref{fig:heuristics_mnt}, whereas the coverage probabilities of the proposed prediction
intervals are presented in fig. \ref{fig:predint_mnt}. Furthermore, simulated average interval
borders and the corresponding true quantiles are given in fig. \ref{fig:hcl_vs_quantile_mnt} 
for three simulation settings, that reflect different levels of possible
skewness of the underlying distribution. 
In these simulation settings H was set to 100 historical control groups. Thus, the historical
information is high enough, that obvious flaws of the methods should be clearly visible and 
not be caused by too low historical sample size.
None of the heuristics control the central 95 \% of the underlying distribution, whereas all 
prediction intervals properly converge to the true quantiles, regardless if the underlying 
distribution is skewed or not.

As expected, the historical range (fig. \ref{fig:heuristics_mnt}A) covers
the simulated future observations $y^*_s$ in all the cases, if the amount of historical control
groups is high enough and lacks a proper control of the statistical error (here 5 \%). Therefore,
the application of the historical range for the calculation of historical control limits should be avoided.

If overdispersion is practically absent ($\phi=1.001$), the control limits of the np-chart 
(fig. \ref{fig:heuristics_mnt}B) yield coverage probabilities relatively close to the nominal 
level. But, with increasing overdispersion, their coverage probabilities decrease below 0.2,
meaning that, by far, less than the desired central 95 \% of the underlying distribution are covered.
This is also demonstrated in fig. \ref{fig:hcl_vs_quantile_mnt}: The average limits of the
np-chart are covered by the true quantiles. Consequently, the np-chart can not be recommended for the application to toxicological (overdispersed) HCD. 

At first glance, it seems that the HCL computed by the mean $\pm$ 2 SD (fig. \ref{fig:heuristics_mnt}C) yield coverage
probabilities close to the nominal level of 0.95, if the amount of historical control 
groups is high enough (at least 20). But, with a decreasing binomial proportion $\pi$ 
and/or an increasing amount of overdispersion $\phi$ the right-skewness of the 
data increases. Hence, in many cases the lower border almost always covers the future observation 
(the triangles in fig. \ref{fig:heuristics_mnt}C approach 1) whereas the coverage 
probability of the upper border (crosses)
does not approach the desired 0.975. This behavior demonstrates that HCL computed
based on the mean $\pm$ 2 SD do not ensure for equal tail probabilities, or in other
words: They do not control the central 95 \% of the underlying distribution,
if this distribution is skewed. This behavior is also demonstrated in fig.
\ref{fig:hcl_vs_quantile_mnt}: The simulated average limits are systematically 
below the corresponding true quantiles, if the underlying distribution is right-skewed.

The coverage probabilities of the prediction intervals proposed in sections 
\ref{sec::bs_pi} and \ref{sec::bayes_pi} are depicted in
Figure \ref{fig:predint_mnt}. 
In general, both frequentist prediction intervals and the
interval based on a Bayesian GLMM yield coverage probabilities relatively close
to the nominal level for most settings. All three intervals behave in a comparable
fashion, except for settings in which the number of historical control groups is
low and / or the amount of overdispersion is high ($H < 10$, $\phi=500$). Contrary to this,
the prediction interval drawn from a Bayesian hierarchical model, fails to control the
statistical error in many of the cases. Furthermore, Bayesian hierarchical models
faced serious convergence issues, if the amount of historical studies and the amount 
of overdispersion was high ($H=100$, $\phi=500$), meaning that the desired interval  
could not be derived for many of the simulated data sets. Consequently, the 
coverage probability could not be computed in this three cases (which explains the 
missing dots for $H=100$ and $\phi=500$ in \ref{fig:predint_mnt}D). 

In the (practical) absence of overdispersion ($\phi=1.01$), the beta-binomial 
prediction interval controls the alpha error best, whereas the prediction intervals
based on the quasi-binomial assumption or Bayesian GLMM tend to behave conservatively, but 
approach the nominal coverage probability with an increase of historical control groups ($H = 100$).
Contrary to this, the prediction interval obtained from Bayesian hierarchical 
modeling remains conservative in this setting, even for 100 historical studies. 

For low to moderate proportions $\pi \in [0.001, 0.1]$ and low to moderate 
overdispersion $\phi \in [3, 50]$ (moderate in the context of $n_h=n^*=18000$),
all intervals, except the one based on Bayesian hierarchical modeling, 
yield coverage probabilities close to the nominal 0.95. 
The interval obtained from Bayesian hierarchical 
modeling approaches the nominal level only for 100 historical studies
and remains liberal for a lower amount of historical information.

The right-skewness of the underlying distribution increases with a decreasing
binomial proportion and an increasing amount of overdispersion 
(panels in the lower left of each sub-graphic). In this case, the 
sampled data contains many zeros and hence, the computed lower borders
tend to always cover the future observation (the triangles approach 1). Hence, 
the prediction intervals practically aim to yield 97.5 \% upper prediction
borders. This effect is also visible in fig.  \ref{fig:hcl_vs_quantile_mnt} 
and reflects a core feature of the underlying distribution,
rather than being based on model misspecification.\\

\newpage

\subsection{Coverage probabilities for the simulation setting inspired by long term carcinogenicity studies}

This part of the simulation was inspired by the estimates for the binomial proportion
and overdispersion reported by Menssen and Schaarschmidt 2019 for HCD obtained from
long term carcinogenicity (LTC) studies run with B6C3F1 mice. 96 combinations of values for the
number of historical studies $H$, the binomial proportion $\pi$ and the 
amount of overdispersion $\phi$ were run, of which 54 combinations directly 
reflect the properties of the reported real life HCD (combinations of the bold faced 
parameter values in tab. \ref{tab::ltc_setting}), whereas the remaining 42 
combinations of parameter values were run in order to enhance the generalisability
of the simulation.

\begin{table}[h!]
\centering
\caption{Parameter values used for the LTC-inspired simulation} 
\begin{tabular}{|c|c|}
\hline 
\textbf{Parameter} & \textbf{Values}$^1$ \\ 
\hline 
No. of historical studies ($H$) & \textbf{5},  \textbf{10},  \textbf{20},  100 \\ 
\hline 
Binomial proportion ($\pi$) & \textbf{0.01},  \textbf{0.1},  \textbf{0.2}, \textbf{0.3}, \textbf{0.4}, \textbf{0.5} \\ 
\hline 
Overdispersion ($\phi$)$^2$ & \textbf{1.001}, \textbf{1.5}, \textbf{3}, 5 \\ 
\hline 
Historical cluster size ($n_h$) & \textbf{50} \\ 
\hline 
Future cluster size ($n^*$) & \textbf{50} \\ 
\hline 
\end{tabular} \\ [0.5ex]
\raggedright
\scriptsize
\textbf{Bold numbers:} Parameter values that reflect estimates obtained from real life HCD.
\textbf{1:} Simulations were run for all 96 combinations of the given parameter values.
\textbf{2:} Since the cluster sizes $n_h$ and $n_*$ are constant with 
$\phi \widehat{=} 1 + (n_h -1) \rho \widehat{=} 1 + (n^* -1) \rho$,
the given values of $\phi$ correspond to intra-class correlation coefficients $\rho$ of 2.04e-05,
0.01020, 0.04081, 0.08163. 
\label{tab::ltc_setting}
\end{table}

Due to the relatively small cluster size, the underlying distribution is 
heavily discrete and can become heavily right skewed for small proportions and/or high
overdispersion (see fig.\ref{fig:hcl_vs_quantile_ltc}). Hence, it is possible that in some settings (especially $\pi=0.01$) 
the data contains many zeros. If it happened that a sampled "historical" data set contained only zeros 
($y_h=0 \quad \forall \quad h=1, \ldots H$), the first observation was set to $y_1=0.5$ and the 
corresponding cluster size was set to $n_1=49.5$ following Menssen and Schaarschmidt 2019. 
This procedure was applied to all frequentist methods (heuristics and prediction intervals) 
in order to enable the estimation process in this extreme
settings, whereas otherwise, the estimates for the binomial proportion and overdispersion 
would become zero. 

For almost all simulation settings, Bayesian hierarchical modeling faced serious 
convergence problems. Therefore, the simulation results are heavily unreliable and not
worth reporting. 
In particular, the precision hyperparameter related to overdispersion is very difficult to 
estimate, which also depends on the prior choice. Therefore, the only Bayesian method 
presented in this section is the prediction interval which is calculated based on a Bayesian GLMM.

\begin{sidewaysfigure}
\centering
 	\includegraphics[width=\textwidth]{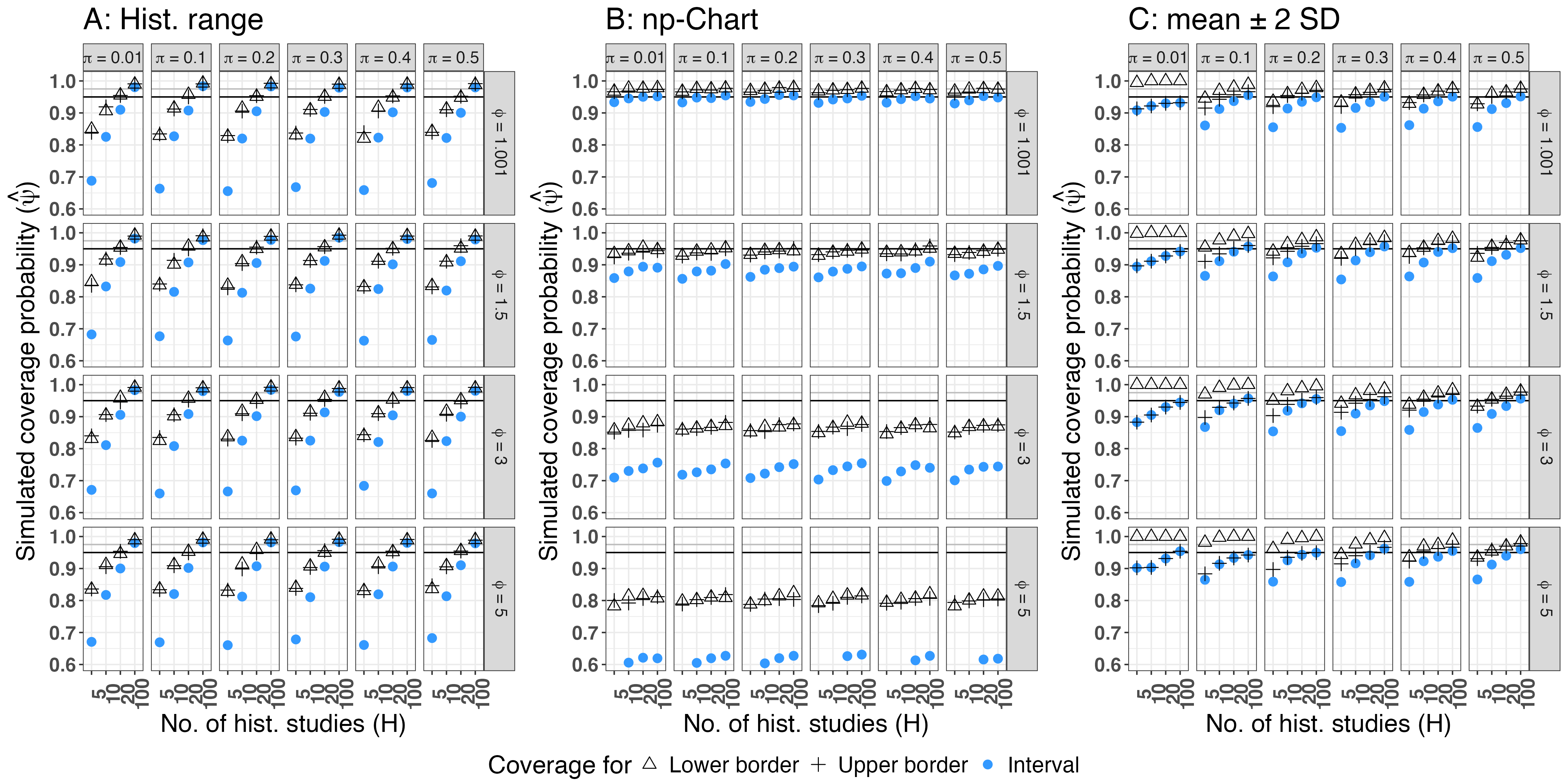}
  \caption{Simulated coverage probabilities of heuristical methods (LTC-setting)
  \textbf{A:} Historical range,
  \textbf{B:} np-chart,
  \textbf{C:} Mean $\pm$ 2 SD,
  \textbf{Black horizontal line:} Nominal coverage probability $\psi^{cp}=0.95$,
  \textbf{Grey horizontal line:} Nominal coverage probability for the lower and 
  the upper limit, if equal tail probabilities are achieved $\psi^{l}=\psi^{u}=0.975$.
  Coverage probabilities below 0.6 are excluded from the graphic}
  \label{fig:heuristics_ltc}
\end{sidewaysfigure}

The heuristical methods (Fig. \ref{fig:heuristics_ltc}) behaved in a similar fashion than
in the MNT-setting: With rising amount of historical control groups, the future observation
is always covered (Fig. \ref{fig:heuristics_ltc}A). 
In the absence of overdispersion, the np-Chart (Fig. \ref{fig:heuristics_ltc}B) yields coverage
probabilities close to the nominal level, but with a rising amount of overdispersion,
its coverage probabilities decrease below 0.6. 
With a rising amount of historical control groups (at least 20), the coverage probabilities 
of the HCL computed by the mean $\pm$ 2 SD (Fig. \ref{fig:heuristics_ltc}C) approach the 
nominal level, if the underlying distribution is relatively symmetrical (low overdispersion and 
relatively high proportion). With increasing right-skewness of the underlying distribution (increasing
overdispersion and decreasing proportion) and /or a low amount of historical control groups ($H<20$)
the HCL computed by the mean $\pm$ 2 SD yield coverage probabilities below the nominal level
that can decrease to approximately 85 \%. 

\begin{sidewaysfigure}
\centering
 	\includegraphics[width=\textwidth]{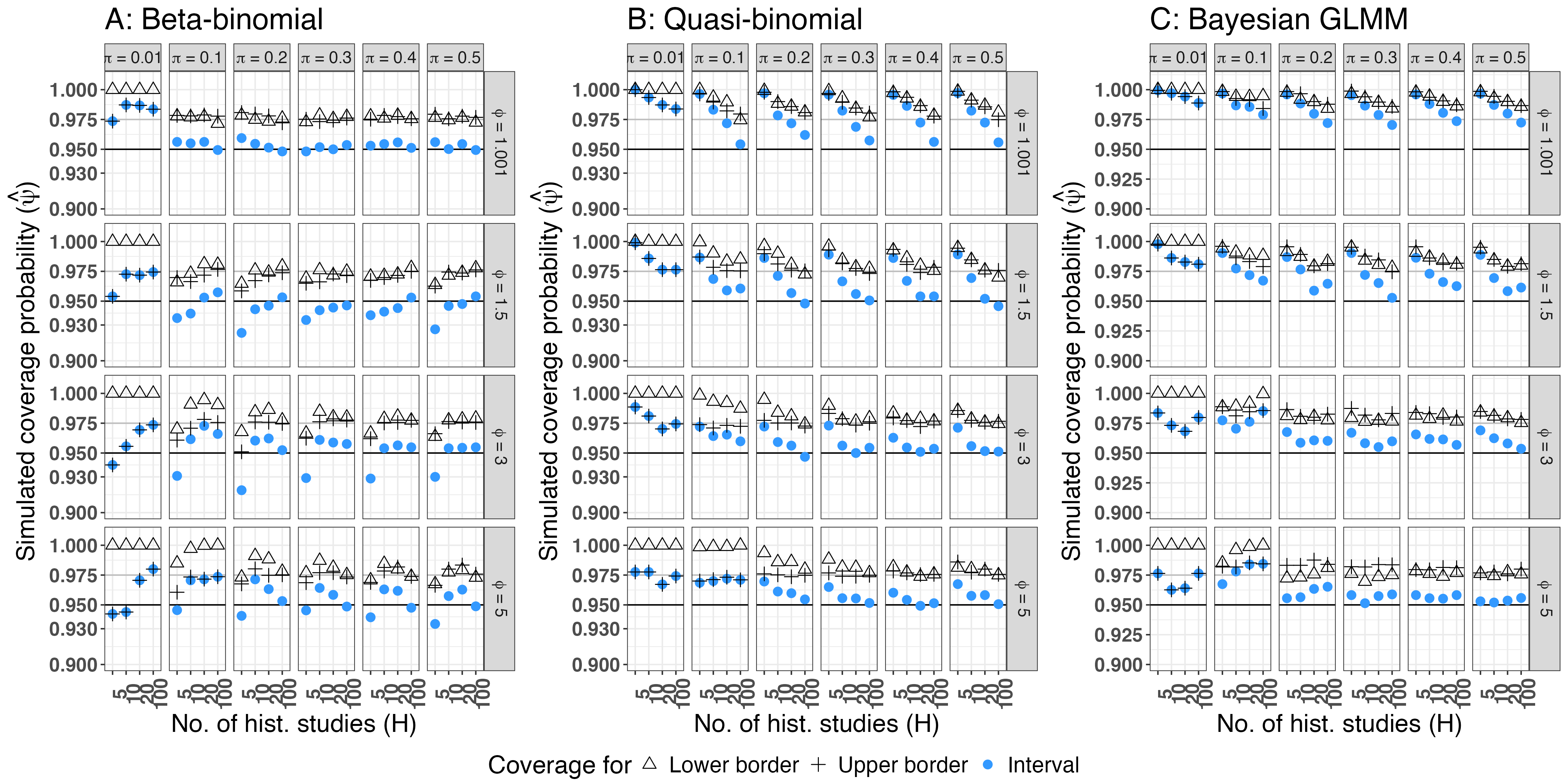}
  \caption{Simulated coverage probabilities of prediction intervals (LTC-setting)
  \textbf{A:} Calibrated beta-binomial prediction interval,
  \textbf{B:} Calibrated quasi-binomial prediction interval,
  \textbf{C:} Prediction interval obtained from a Bayesian GLMM,
  \textbf{Black horizontal line:} Nominal coverage probability $\psi^{cp}=0.95$,
  \textbf{Grey horizontal line:} Nominal coverage probability for the lower and 
  the upper limit, if equal tail probabilities are achieved $\psi^{l}=\psi^{u}=0.975$.}
  \label{fig:predint_ltc}
\end{sidewaysfigure}

\begin{figure}[H]
  \includegraphics[width=\linewidth]{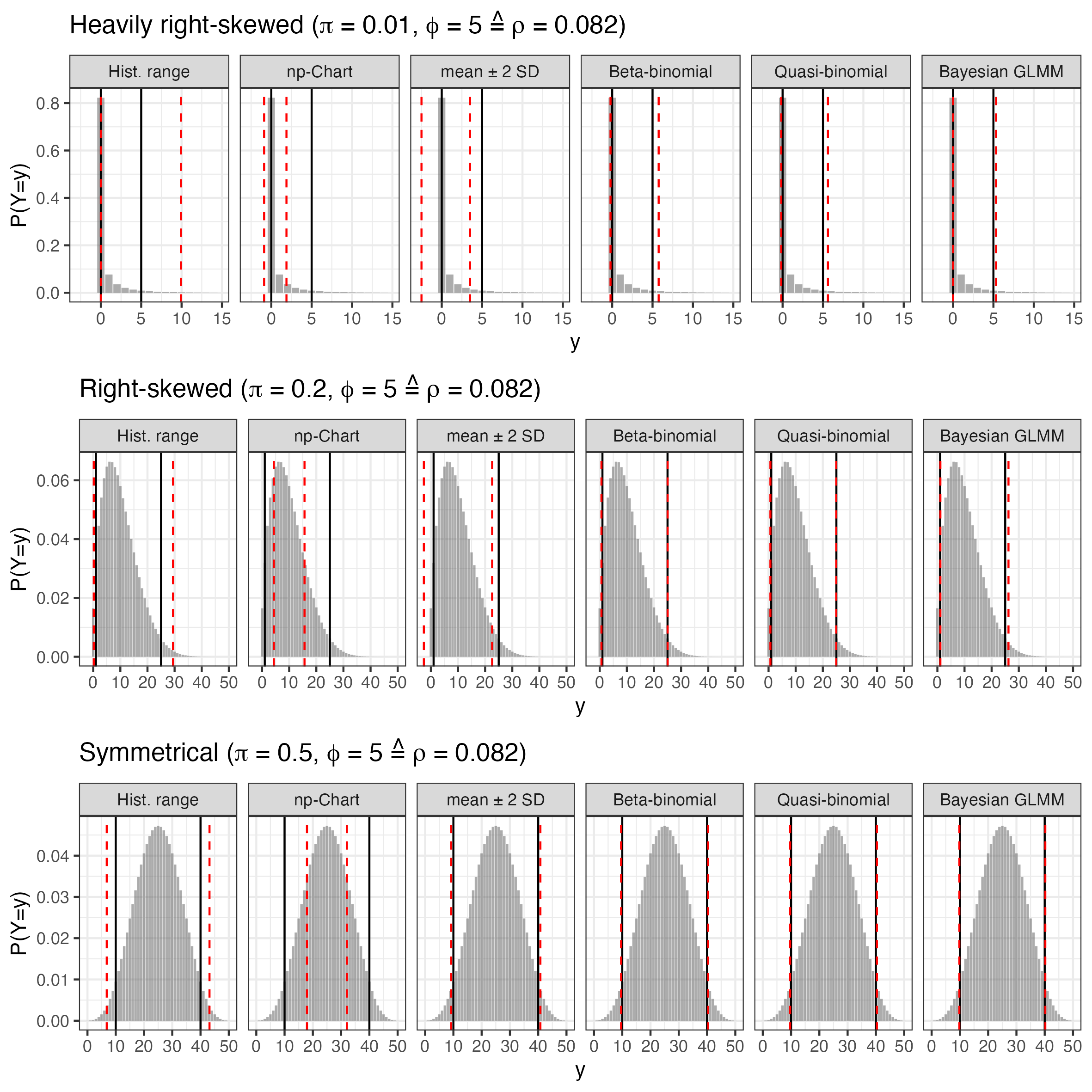}
  \caption{Average limits vs. true quantiles of the underlying distribution (LTC-setting).
  Grey area: Probability mass function of the underlying beta binomial distribution. 
  Black lines: True underlying 2.5 \% and 97.5\% quantiles.
  Dashed red lines: Average limits obtained from the simulation.}
  \label{fig:hcl_vs_quantile_ltc}
\end{figure}

In the practical absence of overdispersion ($\phi=1.001$), the beta-binomial prediction interval (Fig. \ref{fig:predint_ltc}A)
yields coverage probabilities satisfactorily close to the nominal 0.95 (except for $\pi=0.01$), whereas the prediction
intervals that are based on the quasi-binomial assumption (Fig. \ref{fig:predint_ltc}B) or drawn from a Bayesian GLMM 
(Fig. \ref{fig:predint_ltc}C) tend to remain conservative.\\
For $\pi\geq 0.2$, moderate overdispersion ($\phi \in [1.5, 3]$) and at least 10 historical control groups ($H\geq 10$)
the coverage probabilities of the two frequentist prediction intervals satisfactorily approach the nominal 0.95, whereas
the prediction interval obtained from the Bayesian GLMM remains slightly conservative, 
but may also be practically applicable in this scenario. Contrary, for high overdispersion 
($\phi=5$), the quasi-binomial and the Bayesian prediction interval yield coverage probabilities satisfactorily close
to the nominal level (if $\pi\geq 0.2$ and $H\geq 10$). In this setting, the beta-binomial prediction interval yields
coverage probabilities below the nominal level for $H=5$ and remains conservative for $H={10, 20}$, but approaches 
the nominal level for $H=100$. 

With increasing overdispersion $\phi$ and deceasing proportions $\pi\leq 0.1$ the 
right skewness of the underlying distribution increases (see supplementary fig. 1 and 2) and 
its lower 2.5\% quantile can approach 
zero (see fig. \ref{fig:hcl_vs_quantile_ltc}). With other words: in this case the lower prediction
border aims to approximate a true quantile that is zero or close to zero.\\
This is the reason why all three prediction intervals practically yield 97.5 \% upper prediction 
borders for $\pi=0.01$ and the conservativeness of the lower borders increases for $\pi=0.1$ with 
an increase of $\phi$. This is not a misbehavior of the methodologies, but rather reflects a core 
feature of the underlying distribution and one should think about the application of an upper
prediction border, rather than a prediction interval if many zeros are present in the data.


\section{Application of control limits} \label{sec::application}

The calculation of HCL based on real life data is demonstrated using HCD about the mortality 
of male B6C3F1-mice in long-term carcinogenicity studies ($n_h = n^* = 50$). The data set
was provided by Menssen and Schaarschmidt 2019 and contains negative controls from 10 different studies
run between 2003 and 2011 on behalf of the NTP.  It is available via the \texttt{predint}
R package (Menssen 2024) under the name \texttt{mortality\_HCD}. 

The estimate for the binomial proportion is $\hat{\pi}=0.276$ and slight overdispersion of 
$\hat{\phi}=1.31$ appears to be present in the data (estimated with \texttt{stats::glm()}).
The R code for the computation of the  different HCL depicted in table \ref{tab::control_limits_mice} 
is available from GitHub (see link provided in section \ref{sec::comp_det}).

\begin{table}[h!]
\centering
\caption{Historical control limits for the mortality of male B6C3F1-mice obtained in negative controls
of long term carcinogenicity studies at the National Toxicology Program.} 
\begin{tabular}{lrrr}
  \hline
Method & Lower CL & Upper CL & Interval width\\ 
  \hline
  Hist. range & 10 (10) & 21 (21) & 11 (11) \\ 
  np-chart & 7.47 (8) & 20.12 (20) & 12.65 (12) \\ 
  Mean $\pm$ 2 SD & 6.57 (7) & 21.03 (21) & 14.46 (14)\\ 
  Beta-binomial$^1$ & 6.33 (7) & 22.24 (22) & 15.91 (15) \\ 
  Quasi-binomial (QP)$^2$ & 5.77 (6) & 22.71 (22) & 16.94 (16) \\ 
  Bayesian hierarchical$^3$ & 7 (7) &  21.025 (21) & 14.025 (14) \\ 
  Bayesian GLMM & 6 (6) &  23 (23) & 17 (17) \\ 
   \hline
\end{tabular} \\ [0.5ex]
\raggedright
\scriptsize
\textbf{Numbers in brackets:} Lowest and highest number of events covered by the interval.\\
\textbf{1:} Estimates $\hat{\pi}=0.276$ and $\hat{\rho}=0.00621$ were obtained following Lui et al. 2000 (see computational details)\\
\textbf{2:} Estimates $\hat{\pi}=0.276$ and $\hat{\phi}=1.31$ were obtained based on \texttt{stats::glm} (see computational details)\\
\textbf{3:} A weakly informative prior $\kappa \sim \text{Ga}(2,\: 5\cdot 10^{-3})$ was used.
\label{tab::control_limits_mice}
\end{table}
The historical range yields the highest lower limit and the lowest upper limit and hence,
the shortest control interval of all seven methods. Of the six methods that are aimed
to cover a future observation with a probability of 0.95, the np-chart yields the shortest
interval, because it does not account for the presence of overdispersion.\\
Practically, the mean $\pm$ 2 SD and the prediction interval obtained from Bayesian hierarchical 
modeling yield the same limits [7, 21]. Both calibrated prediction intervals practically yield 
upper limits of 22 but the lower limit of the quasi-binomial prediction interval is slightly
lower than the beta-binomial one (6 vs. 7). With limits [6, 23], the prediction interval drawn from
a Bayesian GLMM is the widest one and hence is in line with the simulation results that revealed a 
slightly conservative behaviour for a comparable LTC setting of $\phi=1.5$, $H=10$ and $\pi=0.2$
(see Fig. \ref{fig:predint_ltc}C).\\
Note that, contrary to the description of the Bayesian hierarchical model given in section \ref{sec::comp_det}
(which basically aims at the MNT-inspired setting) a weakly informative prior 
$\kappa \sim \text{Ga}(2,\: 5\cdot 10^{-3})$ was used for the precision hyperparameter $\kappa$. 
This adaption was necessary to reach convergence of the model.


\section{Discussion} \label{sec:discussion}

The validation of a current control group based on prediction intervals is
recommended by several authors (Menssen 2023, Kluxen et al. 2021) and since shortly,
seems to be preferred by the European Food Safety Agency (EFSA 2024) over other (heuristical)
methods. 
Our simulations have shown that heuristical methods do not control the 
statistical error to a satisfactory level and hence, can not be recommended
for the practical validation of a current control group.
Contrary to this, we could show, that three of the four proposed prediction intervals 
satisfactorily control the statistical error, if calculated based on a model 
that match the statistical properties of the HCD. 

\subsection{Bayesian prediction intervals}
The prediction intervals drawn from Bayesian hierarchical models yield coverage probabilities 
that mainly remain below the nominal level in the MNT-setting (if overdispersion is present), 
but the model did not converge 
to most of the simulated data sets that mimic HCD from carcinogenicity studies (mainly if
many zeros were present in the data). It is noteworthy, 
that in the Monte-Carlo simulations, Bayesian hierarchical models were fit, always
using the same prior $\kappa \sim \text{Ga}(2,\: 5\cdot 10^{-5})$ regardless of the 
simulation setting. But, from the application to the real life example
it is obvious, that this type of model needs a case to case adaption of the prior distribution
to provide a reasonable domain for the hyperparameter $\kappa$ (which depends on the amount of overdispersion). 
This might, at least partly,
explain the liberality of this method. Unfortunately, the necessity for a case by case adaption 
of the prior distribution is far beyond the scope of most practitioners who are in charge of the 
reporting of HCD (mainly toxicologists that are barely trained in statistics). Hence this necessity
might be a drawback for practical application.

Contrary to the Bayesian hierarchical models, the Bayesian GLMM did not have convergence problems in the
LTC-setting. But, the applied GLMM mainly yield prediction intervals with coverage probabilities above the nominal level 
(except for a higher amount of overdispersion).
This is, because such models a priori assume between-study 
variability to be present, and hence provide estimates $> 0$ for the respective parameters, even if between-study variability is absent
in the the data-generating process. 

\subsection{Bootstrap calibrated prediction intervals}
The coverage probability of the calibrated intervals is mainly effected by the amount of available historical information, defined by the numbers of experimental units within each control group $n_h$ and number of historical studies $H$. In most of the cases the amount of overdispersion or the magnitude of the 
binomial proportion has minor impact on the coverage probability, as long as one accepts that the 
lower border can become zero for low proportions and / or high overdispersion such that the desired 
95 \% prediction interval practically results in a 97.5 \% upper prediction limit.

For a high cluster size $n_h$ (as in the MNT-setting)
both calibrated prediction intervals yield coverage probabilities satisfactorily close to the
nominal level, even for only five historical control groups. Contrary, for smaller cluster sizes
(as in the LTC-setting) the number of historical control groups has to be slightly higher (say 10) 
to enable coverage probabilities close enough to the nominal level. This difference can be explained 
by the fact, that with increasing numbers of experimental units within each control group ($n_h \to \infty$) the available information on the success rates $\pi_h$ rises. Consequently also their 
estimates $\hat{\pi}_h$ become
more precise. With other words, the precision of $\hat{\pi}_h$ is practically high enough to 
properly estimate the magnitude of the between-study overdispersion if $n_h=18000$ even for $H=5$.
Contrary, if $n_h=50$ (or even lower), the information needed for the estimation of $\hat{\pi}_h$ carried by the data
is heavily limited. Especially if the true (but unknown) success rate $\pi$ 
also gets low, the data contains many zeros. This leads to two implications:

First, the estimates $\hat{\pi}_h$ become heavily imprecise. Second, the estimates for the 
quasi-binomial dispersion parameter $\phi$ as well as for the intra-class correlation $\rho$ are 
known to be negatively biased (McCullagh and Nelder 1989, Saha and Paul 2005).
Consequently, it can happen, that the data practically appears to be underdispersed, even 
if it derives from a data generating process that is inherent of between-study variation
(especially, if most of the $\hat{\pi}_h$ are estimated to be zero). Since underdispersion is biologically
implausible (as explained in the last paragraph of section \ref{sec:Model}), it was necessary to restrict the estimate for the quasi binomial dispersion parameter 
to $max(\hat{\pi}, 1.001)$. Similarly, the estimate for the beta binomial intra-class correlation 
was restricted to be $max(\hat{\rho}, 0.00001)$. Note that both restricted estimates have comparable magnitudes 
of relative bias, but the estimate for the intra-class correlation has a higher coefficient of variation 
(especially if the number of available control groups $H$ is low, see section 6 of the supplementary material). 
These higher variability might explain, that the beta-binomial prediction interval tends to be slightly liberal 
for a low number of 
historical control groups ($H=5$), whereas the quasi-binomial prediction interval behaves in a conservative 
fashion for this scenario.

\subsection{Practical handling of overdispersed data}
The magnitude of overdispersion reflects the between-study variation in toxicological HCD.
However, systematic between-study variation might be induced by a
mixture of both, controllable causes such as different handling of experimental units
by different lab personnel, as well as uncontrollable causes such as the genetic 
variation inherent in the population of experimental units. This means that in practice, overdispersed 
HCD should be handled with care in a way that the magnitude of overdispersion reflects 
mainly the biological variability inherent of the biological process under observation,
rather than the variability that is induced by controllable causes such as different handling 
of experimental units by different lab personnel or the change of the supplier of the model organism
(Dertinger et al. 2023). However, the acceptable magnitude of between-study overdispersion 
depends on the biology of the specific model organism used for a specific type of study. 
Consequently, a general advise on its acceptable magnitude can not be given here, but 
rather has to be discussed separately for each type of study.

\subsection{Perspectives}
The bootstrap calibration procedure presented above was already applied to enable the calculation
of prediction intervals drawn from linear random effects models (Menssen and Schaarschmdit 2022)
or in order to predict overdispersed count data (Menssen et al. 2024). Its further application to 
ovderdispersed binomial data proved, that our algorithm is a general and flexible tool for 
the calculation of prediction intervals. Hence, for the future
it is planned to exploit its potential to enable the calculation of prediction intervals 
for a broader range of models / distributions. Furthermore, it is possible to extend the algorithm to  
different types of simultaneous prediction, which will be the subject of future research.

A highly relevant application not yet covered by this manuscript is the evaluation of HCD on
the base of individual (non-aggregated) values (e.g., for the number of micronuclei for each of
the six individual wells in one historical MNT-experiment rather than for the sum of micronuclei 
per plate) where extra-binomial variability is also possible within one experiment (e.g. between 
the wells of the same plate). For this type of data, generalized linear mixed models need to 
be applied. Jeske and Young 2013 provided a bootstrap based procedure for the calculation of 
prediction intervals based on one-way GLMMs, but to the authors knowledge, a general solution 
for the calculation of prediction intervals drawn from GLMMs is not yet available. In the 
current manuscript we showed how to derive prediction intervals from a Bayesian GLMM with
one level of hierarchy. However, this approach can easily be extended to more complex designs,
but these extensions will need further investigations.


\section{Conclusions} \label{sec::conclusions}

If overdispersion is present in the data (and its magnitude is tolerable):

\begin{itemize}
		\item Commonly applied heuristics (e.g. hist. range, np-charts or mean $\pm$ 2 SD) 
		do not satisfactorily control the statistical error and hence, should not
		be applied for the calculation of HCL.
        \item Symmetrical HCL do not account for possible right- (or left-) skewness
        of the data and hence do not ensure for equal tail probabilities.
        \item In most of the cases, the bootstrap calibrated prediction intervals yield coverage probabilities 
        closest to the nominal level and hence, outperform all other methods reviewed in this 
        manuscript. 
        \item Due to their poor statistical properties (coverage probabilities below the nominal level) and 
        the need for a case to case adaption of prior distributions, prediction intervals drawn from
        Bayesian hierarchical models can not be recommended for practical application.
        \item Prediction intervals computed based on Bayesian GLMM have coverage probabilities 
        which are relatively close to the nominal level (in most cases). Because they can be 
        easily adapted to handle more than one level of hierarchy, they are promising candidates
        for the application to HCD that follows a more complex design.
        \item With an increasing amount of overdispersion and decreasing binomial proportion, the 
        amount of zeros in the data increases. If the amount of zeros in the data is high enough,
        the lower border will always cover the future observation and consequently
        prediction intervals (Bayesian or frequentist) practically yield 97.5 \% upper prediction limits. This  
        behavior is not a drawback of the methodology, but rather reflects one of the core features
        of overdispersed binomial data.
        \item Software for the calculation of bootstrap calibrated prediction intervals is 
        publicly available via the R package predint.
\end{itemize}


\section{Acknowledgments}

We have to thank Frank Schaarschmidt for fruitful discussions on prediction intervals
and the impact of overdispersion on the inference drawn from dichotomous data. 
Furthermore, we have to thank all persons involved in the reviewprocess for their efforts 
and their valuable comments in three rounds of comprehensive review.



\section{References}

Bain, L. J., Patel, J. K. (1993). Prediction intervals based on partial observations for some discrete distributions. IEEE Transactions on Reliability, 42(3), 459-463.\\

Bonapersona, V., Hoijtink, H., RELACS Consortium Abbinck M. 4 Baram TZ 5 6 Bolton JL 5 Bordes J. 7 Knop J. 1 Korosi A. 4 Krugers HJ 4 Li JT 8 9 Naninck EFG 4 Reemst K. 4 Ruigrok SR 4 Schmidt MV 7 Umeoka EHL 4 10 Walker CD 11 Wang XD 12 Yam KY 4, Sarabdjitsingh, R. A., Joëls, M. (2021). Increasing the statistical power of animal experiments with historical control data. Nature neuroscience, 24(4), 470-477. \\

Carlus, M., Elies, L., Fouque, M. C., Maliver, P., Schorsch, F. (2013). Historical control data of neoplastic lesions in the Wistar Hannover Rat among eight 2-year carcinogenicity studies. Experimental and Toxicologic Pathology, 65(3), 243-253. \\

de Kort, M., Weber, K., Wimmer, B., Wilutzky, K., Neuenhahn, P., Allingham, P., Leoni, A. L. (2020). Historical control data for hematology parameters obtained from toxicity studies performed on different Wistar rat strains: Acceptable value ranges, definition of severity degrees, and vehicle effects. Toxicology Research and Application, 4, 2397847320931484. \\

Demétrio, C. G., Hinde, J., Moral, R. A. (2014). Models for overdispersed data in entomology. Ecological modelling applied to entomology, 219-259. \\

Dertinger, S. D., Li, D., Beevers, C., Douglas, G. R., Heflich, R. H., Lovell, D. P., Roberts, D.J., Smith, R.,
Uno, Y., Williams, A., Witt, K.L., Zeller A., Zhou, C. (2023). Assessing the quality and making appropriate use of historical negative control data: A report of the International Workshop on Genotoxicity Testing (IWGT). Environmental and Molecular Mutagenesis. \\

EFSA (2010). Scientific Opinion on the re-evaluation of curcumin (E 100) as a food additive. EFSA J., 8, 1679.\\

EFSA (2024): Draft Scientific Opinion on the use and reporting of historical control data for regulatory studies.
Public Consultation PC-0856. \url{https://connect.efsa.europa.eu/RM/s/consultations/publicconsultation2/a0lTk000000Cvs9/pc0856}\\

Fenech, M. (2000). The in vitro micronucleus technique. Mutation Research/Fundamental and Molecular Mechanisms of Mutagenesis, 455(1-2), 81-95. \\

Fong, Y., Rue, H., Wakefield, J. (2010). Bayesian inference for generalized linear mixed models. 
Biostatistics, 11(3), 397-412. \\

Francq, B. G., Lin, D., Hoyer, W. (2019). Confidence, prediction, and tolerance in linear mixed models. Statistics in medicine, 38(30), 5603-5622. \\

Gabry, J., Goodrich, B., Ali, I., Brillemann, S. (2024). Bayesian applied regression modeling via Stan. 
R Package rstanarm, version 2.32.1 \\

Greim, H., Gelbke, H. P., Reuter, U., Thielmann, H. W., Edler, L. (2003). Evaluation of historical control data in carcinogenicity studies. Human \& experimental toxicology, 22(10), 541-549.\\

Hamada, M., Johnson, V., Moore, L. M., Wendelberger, J. (2004). Bayesian prediction intervals and their 
relationship to tolerance intervals. Technometrics, 46(4), 452-459. \\

Hoffman, D. (2003). Negative binomial control limits for count data with extra‐Poisson variation. Pharmaceutical Statistics: The Journal of Applied Statistics in the Pharmaceutical Industry, 2(2), 127-132. \\

Howley, P. P., Gibberd, R. (2003). Using hierarchical models to analyse clinical indicators: a comparison 
of the gamma-Poisson and beta-binomial models. International Journal for Quality in Health Care, 15(4), 319-329. \\

Jeske, D. R., Yang, C. H. (2013). Approximate Prediction Intervals for Generalized Linear Mixed Models Having a Single Random Factor. Statistics Research Letters, 2(4), 85-95. \\

Kitsche, A., Hothorn, L. A., Schaarschmidt, F. (2012). The use of historical controls in estimating simultaneous confidence intervals for comparisons against a concurrent control. Computational Statistics \& Data Analysis, 56(12), 3865-3875. \\

Kluxen, F. M., Weber, K., Strupp, C., Jensen, S. M., Hothorn, L. A., Garcin, J. C., Hofmann, T. (2021). Using historical control data in bioassays for regulatory toxicology. Regulatory Toxicology and Pharmacology, 125, 105024. \\

Krishnamoorthy K., and T. Mathew. (2009). Statistical Tolerance Regions: Theory, Applications, and Computation. John Wiley and Sons, Hoboken.

Krishnamoorthy, K., Peng, J. (2011). Improved closed-form prediction intervals for binomial and Poisson distributions. Journal of Statistical Planning and Inference, 141(5), 1709-1718.\\

Lovell, D. P., Fellows, M., Marchetti, F., Christiansen, J., Elhajouji, A., Hashimoto, K., Kasamotog, S., Lih, Y., 
Masayasui, O., Moorej, M.M., Schulerk, M., Smithl, R., Stankowski Jr.m,L.F., Tanakag, J., Tanirn, J.Y., 
Thybaudo, V., Van Goethemp F., Whitwell, J. (2018). Analysis of negative historical control group data from the in vitro micronucleus assay using TK6 cells. Mutation Research/Genetic Toxicology and Environmental Mutagenesis, 825, 40-50. \\

Lui, K. J., Mayer, J. A., Eckhardt, L. (2000). Confidence intervals for the risk ratio under cluster sampling based on the beta‐binomial model. Statistics in medicine, 19(21), 2933-2942. \\

Levy, D. D., Zeiger, E., Escobar, P. A., Hakura, A., Bas-Jan, M., Kato, M., Moore, M.M., Sugiyama, K. I. (2019). Recommended criteria for the evaluation of bacterial mutagenicity data (Ames test). Mutation Research/Genetic Toxicology and Environmental Mutagenesis, 848, 403074. \\

McCullagh P, Nelder J.A. (1989): Generalized Linear Models. 2nd Edition, Chapman and Hall, London. \\

Meeker, W. Q., Hahn, G. J., Escobar, L. A. (2017). Statistical intervals: a guide for practitioners and researchers. John Wiley \& Sons. \\

Menssen, M., Schaarschmidt, F. (2019). Prediction intervals for overdispersed binomial data with application to historical controls. Statistics in Medicine, 38(14), 2652-2663. \\

Menssen, M.,  Schaarschmidt, F. (2022). Prediction intervals for all of M future observations based on linear random effects models. Statistica Neerlandica, 76(3), 283-308. \\

Menssen, M. (2023). The calculation of historical control limits in toxicology: Do's, don'ts and open issues from a statistical perspective. Mutation Research/Genetic Toxicology and Environmental Mutagenesis, 503695. \\

Menssen M. (2024). predint: Prediction Intervals. R package version 2.2.1 \\

Menssen, M., Dammann, M., Fneish, F., Ellenberger, D.,  Schaarschmid, F. (2024). Prediction intervals for overdispersed Poisson data and their application in medical and pre-clinical quality control. Pharmaceutical Statistics (published online). https://doi.org/10.1002/pst.2447 \\

Millard SP (2013). EnvStats: An R Package for Environmental Statistics. Springer, New York. ISBN 978-1-4614-8455-4 \\

Montgomery, D. C. (2019). Introduction to statistical quality control. John Wiley \& Sons. \\

Nelson, W. (1982). Applied Life Data Analysis. New York, NY: John Wiley \& Sons. \\

NTP (2018): NTP Technical Report 595: Toxicology and Carcinogenesis Studies in Hsd:Sprague Dawley SD Rats Exposed to Whole-body Radio Frequency Radiation at a Frequency (900 MHz) and Modulations (GSM and CDMA) Used by Cell Phones. \url{https://ntp.niehs.nih.gov/publications/reports/tr/tr595}\\

NTP (2024): NTP Historical Controls Data Base. \url{https://ntp.niehs.nih.gov/data/controls} \\

OECD 471: Bacterial Reverse Mutation Test, OECD Guidelines for the Testing of Chemicals, Section 4, OECD Publishing, Paris. \\

OECD 473: In Vitro Mammalian Chromosomal Aberration Test, OECD Guidelines for the Testing of Chemicals, Section 4, OECD Publishing, Paris \\

OECD 487: In Vitro Mammalian Cell Micronucleus Test, OECD Guidelines for the Testing of Chemicals, Section 4, OECD Publishing, Paris. \\

OECD 489: In Vivo Mammalian Alkaline Comet Assay, OECD Guidelines for the Testing of Chemicals, Section 4, OECD Publishing, Paris. \\

OECD 490: In Vitro Mammalian Cell Gene Mutation Tests Using the Thymidine Kinase Gene, OECD Guidelines for the Testing of Chemicals, Section 4, OECD Publishing, Paris. \\

Prato, E., Biandolino, F., Parlapiano, I., Grattagliano, A., Rotolo, F., Buttino, I. (2023). Historical control data of ecotoxicological test with the copepod Tigriopus fulvus. Chemistry and Ecology, 39(8), 881-893. \\

Rotolo, F., Vitiello, V., Pellegrini, D., Carotenuto, Y., Buttino, I. (2021). Historical control data in ecotoxicology: Eight years of tests with the copepod Acartia tonsa. Environmental Pollution, 284, 117468. \\

Ryan, L. (1993). Using historical controls in the analysis of developmental toxicity data. Biometrics, 1126-1135.\\

Spiliotopoulos, D., Koelbert, C., Audebert, M., Barisch, I., Bellet, D., Constans, M.,  
Czich, A., Finot, F., Gervais, V., Khoury, L., Kirchnawy, C., Kitamoto, S., Le Tesson, A.,
Malesic, L., Matsuyama, R., Mayrhofer, E., Mouche, I., Preikschat, B., Prielinger, L., 
Rainer, B., Roblin, C., Wäse, K. (2024). Assessment of the performance of the Ames MPF™ assay:
A multicenter collaborative study with six coded chemicals. Mutation Research/Genetic Toxicology and Environmental Mutagenesis, 893, 503718. \\

Stan Development Team. (2024) RStan: the R interface to Stan. R package version 2.32.6. \\

Stan Development Team (2024). Stan functions reference. version 2.35.27 \\

Tarone, R. E. (1982a). The use of historical control information in testing for a trend in proportions. Biometrics, 215-220. \\

Tarone, R. E. (1982b). The use of historical control information in testing for a trend in Poisson means. Biometrics, 457-462.\\

Tian, Q., Nordman, D. J., Meeker, W. Q. (2022). Methods to compute prediction intervals: A review and new results. Statistical Science, 37(4), 580-597.\\

Tug, T., Duda, J. C., Menssen, M., Bruce, S. W., Bringezu, F., Dammann, M., Frotschl, R., Harm, V., Ickstadt, K.,
Igl, B-W., Jarzombek, M., Kellner, R., Lott, J., Pfuhler, S., Plappert-Helbig, U., Rahnenfuhrer, J., Schulz, M.,
Vaas, L., Vasquez, M., Ziegler, V., Ziemann, C. (2024). In vivo alkaline comet assay: Statistical considerations on historical negative and positive control data. Regulatory Toxicology and Pharmacology, 148, 105583. \\

Valverde-Garcia, P., Springer, T., Kramer, V., Foudoulakis, M.,  Wheeler, J. R. (2018). An avian reproduction study historical control database: a tool for data interpretation. Regulatory Toxicology and Pharmacology, 92, 295-302. \\

Vandenberg, L. N., Prins, G. S., Patisaul, H. B., Zoeller, R. T. (2020). The use and misuse of historical controls in regulatory toxicology: lessons from the CLARITY-BPA study. Endocrinology, 161(5), bqz014. \\

Wang, H. (2008). Coverage probability of prediction intervals for discrete random variables. Computational statistics and data analysis, 53(1), 17-26.\\

Wang, H. (2010). Closed form prediction intervals applied for disease counts. The American Statistician, 64(3), 250-256.\\

WHO (2024). Pesticide residues in food 2022. Joint FAO/WHO meeting on pesticide residues. Evaluation Part II–Toxicological. World Health Organization. \\

Zarn, J. A., König, S. L., Shaw, H. V., Geiser, H. C. (2024). An analysis of the use of historical control data in the assessment of regulatory pesticide toxicity studies. Regulatory Toxicology and Pharmacology, 105724.

\end{document}